# Multi-Head Attention Neural Network for Smartphone Invariant Indoor Localization

Saideep Tiku, Danish Gufran, Sudeep Pasricha
Department of Electrical and Computer Engineering
Colorado State University, Fort Collins, CO, United States
{saideep, danish.gufran, sudeep}@colostate.edu

*Abstract*—Smartphones together with RSSI fingerprinting serve as an efficient approach for delivering a low-cost and high-accuracy indoor localization solution. However, a few critical challenges have prevented the wide-spread proliferation of this technology in the public domain. One such critical challenge is device heterogeneity, i.e., the variation in the RSSI signal characteristics captured across different smartphone devices. In the real-world, the smartphones or IoT devices used to capture RSSI fingerprints typically vary across users of an indoor localization service. Conventional indoor localization solutions may not be able to cope with device-induced variations which can degrade their localization accuracy. We propose a multi-head attention neural network-based indoor localization framework that is resilient to device heterogeneity. An in-depth analysis of our proposed framework across a variety of indoor environments demonstrates up to 35% accuracy improvement compared to state-of-the-art indoor localization techniques.

## I. INTRODUCTION

The proliferation of GPS (Global Positioning Systems) technology revolutionized the way we navigate and interact with the world around us. Today, every smartphone comes with a built-in GPS which is invaluable for outdoor navigation. The same technology is now powering several geo-location-based businesses and other interactive platforms such as Uber, Pokémon-Go and location-based marketing. Today, indoor localization technology holds a similar potential to disrupt the way we navigate within spaces that are unreachable by GPS, e.g., malls, buildings, and tunnels. Several startups such as IndoorAtlas, Target (Shopkick), and Zebra have already started to provide services that can help customers find products within a store [1].

Unlike GPS for outdoor localization, there are no generalized global standards for indoor localization. Therefore, a myriad of techniques have been developed that use a variety of sensors and radio frequencies. Some commonly utilized radio signals for indoor localization include Bluetooth, RFID, UWB (Ultra-Wide Band), and WiFi [2]-[4]. Among these, WiFi based indoor localization has been the most widely researched, due to its low setup costs and easy availability in almost all indoor locales [3]. The ubiquity of WiFi based indoor localization is further fueled by the fact that most people today own smartphones with WiFi radios.

Despite the advantages of WiFi based indoor localization, there are also some open challenges. WiFi signals suffer from weak wall penetration, multipath fading, and shadowing effects. These challenges make it difficult to establish a direct mathematical relationship between Received Signal Strength Indicator (RSSI) and distance from WiFi Access Points (APs). These issues have served as a motivation to use fingerprinting-based techniques. Fingerprinting is based on the idea that different locations indoors exhibit a unique signature of AP RSSI values. Due to its independence from the RSSI-distance relationship, fingerprinting overcomes some of the aforementioned drawbacks of WiFi based indoor localization.

Traditionally, fingerprinting-based indoor localization consists of two phases. In the *first* phase (offline or training phase), the provider of the localization service captures the RSSI values for visible APs at various indoor locations of interest. This results in a database of RSSI fingerprint vectors and associated locations or Reference Points (RPs). This database may further be used to train models (e.g., machine learning-based) for location estimation. In the *second* phase (online or testing phase), a user unaware of their own location captures an RSSI fingerprint on a device such as a smartphone. This fingerprint is then sent to the trained model to determine the user's location. This location can be overlaid on a map and visualized on the smartphone's display.

A vast majority of work in the domain of fingerprinting-based indoor localization employs the same smartphone for (offline) data collection and (online) location prediction, e.g., [5]-[7]. This approach assumes that in a real-world setting, the localization devices operated by users would have identical signal characteristics. In today's diverse smartphone market, consisting of various brands and models, such an assumption is largely invalidated. In reality, the smartphone user base is comprised of heterogeneous mobile devices that vary in antenna gain, WiFi chipset, antenna shape, OS version, etc. Recent works have shown that the perceived RSSI values for a given location captured by different smartphones can vary significantly [8]. This variation degrades the localization accuracy achieved through conventional fingerprinting. Therefore, there is a need for smartphone heterogeneity invariant (i.e., device invariant) fingerprinting techniques.

In this paper, we present a robust and computationally lightweight WiFi RSSI-based fingerprinting framework (*ANVIL*) that aims to achieve device invariance such that it experiences minimal accuracy loss across heterogeneous smartphones. The main contributions of our work are:

- we conduct an analysis to highlight the impact of device heterogeneity on RSSI fingerprints;

- towards the goal of promoting generalized device invariance, we identify and adapt data augmentation methodologies for training deep machine learning models;

- we introduce a novel multi-head attention neural network for device invariant indoor localization;

- we create a set of benchmarks by collecting fingerprints with multiple heterogeneous devices across buildings, to test the localization accuracy of *ANVIL* against state-of-the-art indoor localization techniques;

- we prototype our calibration-free device invariant indoor localization framework, deploy it on smartphones, and evaluate it under real-world settings.

## II. Related Work

A considerable amount of work has been dedicated to addressing the challenges associated with WiFi fingerprinting based indoor localization. The recent growth in the computation capabilities of smartphones has enabled the proliferation of localization algorithms and frameworks with high computational and memory demands. For instance, Feed-Forward Deep Neural Networks (FF-DNN) [9] and also more sophisticated Convolutional Neural Networks (CNN) [6] combined with ensemble learning are being used in smartphones to improve indoor localization accuracy [7]. One of the concerns with utilizing such techniques is the severe energy limitations on mobile devices. Pasricha et al. [5] proposed an energy-efficient fingerprinting-based technique. However, most prior works in the domain of fingerprinting-based indoor localization, including [5], are plagued by the same major drawback, i.e., the lack of an ability to adapt to device heterogeneity across the offline and online phases. This drawback, as discussed in section VI, leads to unpredictable degradation in localization accuracy in the online phase.

In the offline phase, the localization service provider likely captures fingerprints using a device that is different from the IoT devices or smartphones employed in the online phase by users. Some common software and hardware differences that introduce device heterogeneity include WiFi antennas, smartphone design materials, hardware drivers, and the OS [4]. Techniques to overcome such issues fall into two major categories: *calibration-based* and *calibration-free* methods.

The calibration of an indoor localization framework can take place in either the offline or online phase. The work in [10] employs offline phase calibrated fingerprinting-based indoor localization. In this approach, the fingerprint collection process employs a diverse set of devices. Later, an autoencoder is specifically trained and calibrated using fingerprints from different devices. The role of the autoencoder is to create an encoded latent representation of the input fingerprint from one device such that decoded output fingerprint belongs to another device. In this manner, the latent representation created is expected to be device invariant. While this approach is promising, it comes with the significant overhead of utilizing multiple devices in the offline phase to capture fingerprints. Additionally, there are no guarantees that the set of devices employed to capture fingerprints in the offline phase capture ample heterogeneity that may be experienced by the localization framework in the online phase.

The online phase calibration approach involves acquiring RSSI values and location data manually for each new device in the online phase [11]. Such an approach is however not very practical as it can be cumbersome for users. With this approach, once a user arrives at an indoor locale, they have to move to a known location and capture an RSSI fingerprint. The RSSI information is collected, and then manually calibrated through transformations such as weighted-least square optimizations and time-space sampling [12]. These techniques can be aided by crowdsourcing schemes. However, such approaches suffer from accuracy degradation [13].

In calibration-free fingerprinting, the fingerprint data is often translated into a standardized form that is portable across devices. One such approach, known as Hyperbolic Location fingerprint (HLF) [14], uses the ratios of individual AP RSSI values to form the fingerprint. Unfortunately, HLF significantly increases the dimensionality of the training data in the offline phase. The Signal Strength Difference (SSD) approach [15] reduces the dimensionality by taking only independent pairs of WiFi APs into consideration. But this approach suffers from low accuracy. An alternative to the standardization of fingerprints is presented in [16]. The work proposed AdTrain, which employs adversarial training to improve the robustness of a deep-learning model against device heterogeneity. AdTrain introduces noise in the fingerprint and associated location label before training the deep-learning model. Based on our experiments, such an approach does improve the deep-learning model's robustness to device heterogeneity in a limited manner. The approach used in our framework in this paper (*ANVIL*) is orthogonal to the one proposed in AdTrain and *ANVIL* can be extended to include AdTrain. The work in [17] employs Stacked Auto-Encoders (SAE) for improving resilience to device heterogeneity. The authors expect that the lower dimensional encodings created using the SAE are more resilient to RSSI variations across devices. However, based on our experimental evaluations (section VI.B), we found that such an approach is unable to deliver high-quality localization and also does not converge easily. In contrast, *ANVIL* employs a multi-head attention neural network to achieve improvements in device invariance for indoor localization.

There are a limited set of previous works that have employed attention layers for magnetic [18], [19] and Channel State Information (CSI) fingerprint-based [21] indoor localization. However, these works do not consider device heterogeneity. Additionally, it is well known that magnetic fingerprints are unstable over time and heavily impacted by minor changes in the indoor environment. On the other hand, CSI enabled hardware is not available in off-the-shelf smartphones available today [20]. In contrast, *ANVIL* employs relatively stable WiFi RSSI that can be captured through heterogeneous off-the-shelf smartphones of various vendors to deliver stable indoor localization.

Our proposed framework in this paper, *ANVIL*, is a multi-head attention neural network based and calibration-free indoor localization framework that can be easily deployed to off-the-shelf smartphones, while relying on the ubiquity of WiFi signals to deliver device invariant performance. *To the best of our knowledge, no previous works have attempted to use attention in the context of improving device invariance.* We performed extensive evaluation of *ANVIL* against state-of-the-art prior works and evaluated it over a variety of indoor environments and smartphones in real-world settings.

## III. Analysis of Heterogeneous Fingerprints

To understand the cause of degradation in localization performance due to device heterogeneity, we evaluate the RSSI fingerprints captured by two distinct smartphones. For this experiment, 10 fingerprints are captured at a single location with two smartphones: LG V20 (LG) and Oneplus 3 (OP3). The RSSI values captured are in the range of –100dB to 0dB, where –100dB indicates no received signal and 0dB is the highest signal strength. The RSSI values for the two devices are presented in figure 1. The solid lines represent the mean values, whereas the shaded regions represent the range of the observed RSSI values. From figure 1, we can make the following key observations:

- There is considerable similarity in the shape of the RSSI fingerprints across the two smartphones. Therefore, indoor localization frameworks that focus on pattern-matching-based approaches may be able to deliver higher quality localization performance.

- The RSSI values for the LG device exhibit an upward shift (higher signal reception) by an almost constant amount.

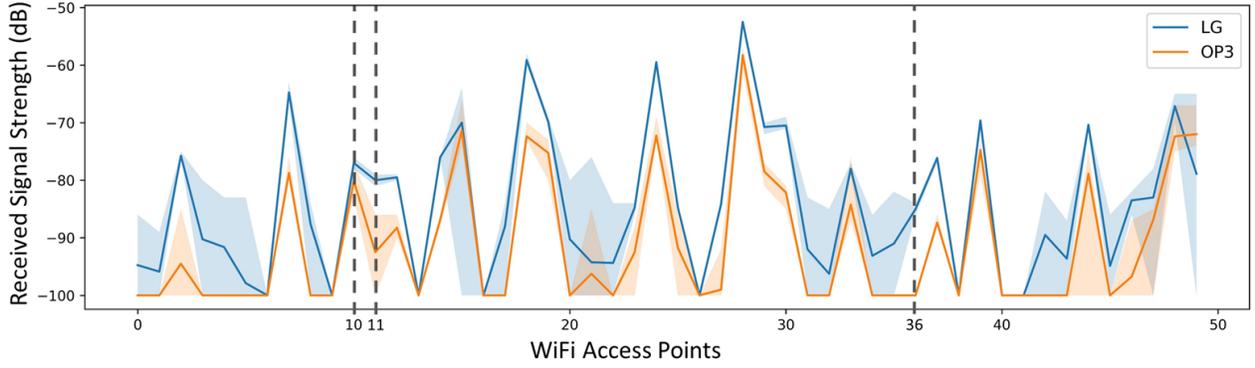

Figure 1. RSSI values for WiFi APs as observed by two different smartphones at a specific location. The solid lines represent the mean values. Shaded regions represent the range of RSSI values, over 10 fingerprint readings. Distinct WiFi APs (by MAC IDs) are denoted as unique integers from 0 to 50.

This constant shift of RSSI value is similar to increasing the *brightness* of an image.

- We also observe a *contrastive* effect for certain parts of the fingerprints. By contrastive effect, we mean the difference between the RSSI value changes. In figure 1, we observe that while the RSSI fingerprints across the two devices rise and fall together, the specific amount that they rise and fall by are not the same. A clear example of this observation can be seen across the RSSI values of APs 10 and 11.

- Some APs that are visible when using the LG device are never visible (RSSI = –100dB) to the OP3 device. A good example of this is AP 36, where the RSSI value for the LG device is –80dB and the RSSI value for the OP3 device always remains –100dB. From the perspective of deep-learning models, this is similar to the random *dropout* of input AP RSSI.

- Lastly, we also observe that the RSSI values from the LG device vary much more than the OP3 device, for a majority of the APs. This indicates that different smartphones may experience different amounts of variation in RSSI values, and it may also be hard to quantify the range of RSSI variation for a given device beforehand. Therefore, our proposed approach must be resilient to such unpredictable variations in the raw RSSI values observed from an AP.

The evaluation of observed fingerprint variations across the two smartphones presented in this section (as well as our analysis with other smartphones that show a similar trend) guides the design of our device invariant *ANVIL* framework. We specifically base our data augmentation strategy in *ANVIL* on the observations made here. The multi head attention-based deep-learning approach (discussed in the next section) was chosen and adapted in *ANVIL* to promote generalized pattern matching across heterogeneous devices.

## IV. THE ATTENTION MECHANISM

The attention mechanism in the domain of deep learning is a relatively new approach initially employed for natural language processing (NLP) [22] and more recently also for image classification [23]. The concept of attention is derived from the idea of human cognitive attention and our ability to selectively focus on sub-components of information while ignoring less relevant components of the same. For the purpose of deep learning, the attention mechanism is modeled as the retrieval of attention information from a database containing key and value pairs. Given a query $Q$ and a set of key-value pairs $(K, V)$, attention can be computed as:

$$Attention(Q, K, V) = similarity(Q, K)V \quad (1)$$

In the above expression, a variety of similarity functions may be employed such as dot-product, scaled dot product, additive dot product, etc. [18]. Among the variants, the dot-product attention is the fastest to compute and is the most space-efficient, making it our choice for this work.

The process of computing scaled-dot-product attention (figure 2; right) for a given set of queries, keys, and values $(Q, K, V)$ can be captured in an equation as:

$$Attention(Q, K, V) = softmax\left(\frac{Q K^T}{\sqrt{d_k}}\right)V \quad (2)$$

where $d_k$ is the dimensionality of the key vector. The role of the scaling factor ($\sqrt{d_k}$) is to counteract the effects of very large magnitudes being fed to the Softmax function, leading to regions that produce extremely small gradients. In practice, scaled dot-products are also known to outperform other approaches of computing attention.

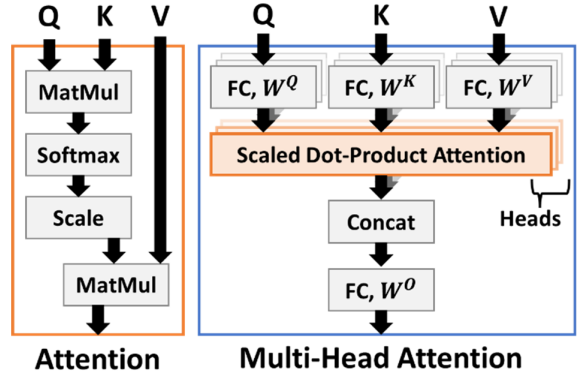

Figure 2. A procedural representation of scaled-doct-product attention (left) and multi-head attention (right). Each attention layer accepts three inputs: Query (Q), Keys (K), and Values (V).

While attention serves as a low-overhead approach for capturing a weighted relationship between queries and values, given a set of key-value pairs, it lacks any learnable parameters. This limits its ability to identify and quantify the hidden relationships between the different pairings of $(Q, K, V)$. The work in [22] extends the idea of a singular attention computation with multiple distinct versions (multiple heads) of linearly projected queries, keys, and values. Each of these learnable linear projections are of dimension $d_k$, $d_q$, and $d_v$. This form of attention is more commonly known as multi-headed attention and can be formally captured by the following equation:

$$MultiHead(Q, K, V) = Concat(h_1, h_2, h_3 \ldots h_n)W^o$$
$$where\ h_i = Attention(QW_i^Q, KW_i^K, VW_i^V) \quad (3)$$

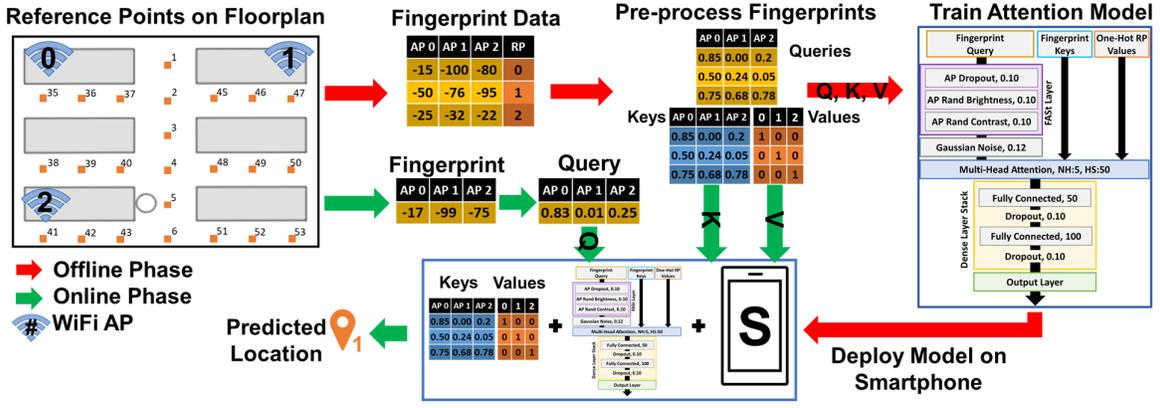

Figure 3. An overview of the ANVIL indoor localization framework depicting the offline (red arrows) and online (green arrows) phases.

where $W_i^Q$, $W_i^K$, $W_i^V$ and $W_i^O$ are model parameters or weights associated with the linear projections with dimensions *head size × d*. Head size (HS) is a hyperparameter of the multi-head attention layer and *d* is the length of vector being projected (query, key or value). The process of computing scaled-dot-product attention and multi-headed attention is depicted in figure 2. The computation associated with each head can be performed in parallel, as shown.

## V. ANVIL FRAMEWORK

### A. Overview

A high-level overview of our proposed *ANVIL* framework is presented in figure 3. We begin in the offline phase (annotated by red arrows), where we first capture RSSI fingerprints for various RPs (see section VI for details on RP granularity) across the floorplan of the building. Each row in the RSSI database consists of the RSSI values for every AP visible across the floorplan and its associated RP. These fingerprints are then pre-processed into queries (*Q*), keys (*K*) and values (*V*) to train a multi-headed attention neural network model, as shown in figure 3. The specific details of fingerprint preprocessing, model design and training are covered later in this section. Once the model has been trained, it is then deployed on a smartphone, along with the fingerprint key-value pairings. This concludes the offline stage.

In the online phase (green arrows), the user captures an RSSI fingerprint vector at an RP that is unknown. For any WiFi AP that was visible in the offline phase and is not observed in this online phase, its RSSI value is assumed to be –100dB, ensuring consistent RSSI vector lengths across the phases. This fingerprint is pre-processed (see Section V.B) to form the fingerprint query and sent to the deployed multi-head attention neural network model on the smartphone. As shown in figure 3, the model when fed the online phase fingerprint query, along with the keys and values from the offline phase, produces the location of the user in the online phase. This location is then shown to the user on the smartphone's display.

In the following subsections, we elaborate on the major components of the *ANVIL* framework depicted in figure 3.

### B. RSSI Fingerprint Preprocessing

The RSSI for various WiFi APs along with their corresponding reference points are captured within a database as shown in figure 3. As mentioned earlier, the RSSI values vary in the range of –100dB to 0dB, where –100 indicates no signal and 0 indicates a full (strongest) signal. The RSSI values captured in this dataset are normalized to a range of 0 to 1, where 0 represents the weak or null signal, and 1 represents the strongest signal. This new dataset is the basis of all training data required by our multi-head attention model. As discussed in Section IV, the multi-head attention model requires three main inputs: the Queries and the Key-Value pair dataset. For this work, the RSSI fingerprint vectors captured in the training phase (without RP information) are used as both queries and keys in the offline phase. The associated RPs are one-hot encoded and are used as values.

### C. Fingerprint Augmentation Stack (FASt) Layer

A major challenge to maintaining localization stability for fingerprinting-based indoor localization is the variation in RSSI fingerprints across heterogeneous devices, as discussed in Section III. However, in the online phase, it would be impossible to foretell what combination of effects the fingerprint captured using a smartphone. The received RSSI fingerprints can vary depending on external factors like building layout, walls, metallic object (reflecting the RSSI fingerprint) making the prediction of the location ambiguous.

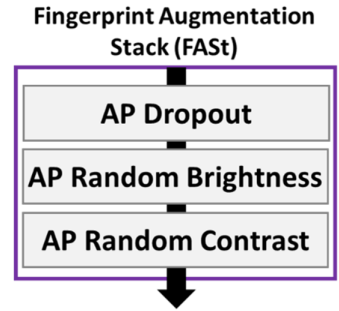

Figure 4. Various components of the Fingerprint Augmentation Stack (FASt). Each subcomponet augments a specific heterogeneity effect.

To overcome this challenge, we propose a streamlined Fingerprint Augmentation Stack (FASt) implemented as a layer that contains the required subcomponents for the augmentation of RSSI fingerprints that would promote the resilience to device heterogeneity. The major advantage of FASt is that it can be seamlessly integrated into any deep-learning based model and promote the rapid prototyping of device invariant models for the purpose of fingerprinting-based indoor localization. A procedural representation of the FASt layer is shown in figure 4.

The design of the FASt layer is based on the three observations from our analysis in Section III, i.e., AP dropout, contrast, and brightness. While these are well known data augmentation techniques in the domain of computer vision [24], they do not directly translate to the domain of RSSI fingerprinting-based pattern matching. In the computer vision domain, all three effects are applied to the inputs indiscriminately, as it is most likely that the input does not contain cropped portions relative to rest of the image (blacked/whited out regions of image). This is especially true in the domain of image-based pattern matching, where the input image may not have cropped out components, with the exception of image inpainting [25]. The utilization of image augmentation before training deep-learning models is a well-known approach for improving generalizability.

In the case of RSSI-fingerprints, the fact that a certain set of APs are not visible in some areas of the floorplan as compared to the others is a critical and unique attribute. A deep-learning framework might utilize this to better correlate an RSSI pattern to a specific location on the floorplan. Based on these beliefs and our analysis in Section III, we adapt the general random dropout, brightness, and contrast layers to only act on WiFi APs that are visible to the smartphone i.e., $RSSI \neq -100\ dB$. It is important to note that we deliberately, do not add random noise to the FASt layer, as it was not specially adapted for the purpose of RSSI fingerprinting-based indoor localization. However, random noise is an important aspect of the model design as discussed in the next sub-section.

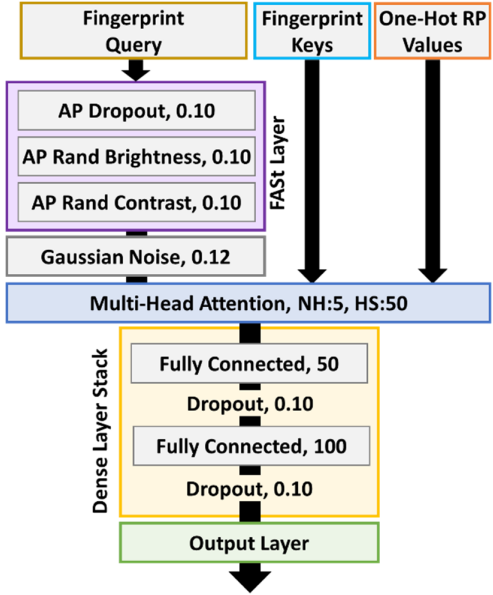

Figure 5. Overview of the multi-head attention neural network model in the *ANVIL* framework.

### D. Multi-Head Attention Model

The concept of attention and especially multi-head has gained considerable popularity in various domains of pattern matching [22]. Its simplicity in mathematical operations enables a computationally efficient process for deep learning. It is also well known that conventional feed-forward neural network layers (dense layers) require lower computational capabilities on the target deployment platform when compared to convolutional approaches. Additionally, recent work in the domain of pattern matching is slowly considering attention as an alternative to convolutional layers [26]. Based on these observations, we combine our domain specific fingerprint augmentation layer stack (FASt), Multi-Head Attention, and feed-forward neural networks to create a deep-learning classification model as depicted in figure 5.

The model shown in figure 5 consists of three inputs: fingerprint query, keys, and one-hot encoded reference point (RP) labels as values. The key-value inputs remain fixed in the offline and online phases of the model development process. The fingerprint keys and RP values are the fingerprints captured at specific RPs using the smartphone from the offline phase. In the offline phase, we use the same fingerprints as queries to train the model, whereas, in the online phase, fingerprints captured by the end-user are fed as queries to the model which in turn produces the location of the user.

For the sake of simplicity, all hyperparameters in the model were chosen such that the model generalizes well across all the floorplans discussed in the experiments section. We employ the value of 0.10 for the dropout, random brightness, and contrast functions of the FASt layer. Therefore, for each RSSI fingerprint Query fed to the model, random dropout, random contrast, and random brightness (increase or decrease) are applied with a 10% probability. Similarly, the Gaussian noise layer was set to a standard deviation of 0.12. No augmentation is applied to the fingerprint keys. Based on our experiments (not presented for brevity), the overall performance of the model is better when Gaussian noise is applied to the whole fingerprint, instead of the masked approach ($RSSI \neq -100$ dB) used for the FASt layer. One possible explanation is that the noise layer acts as regularization on the visible AP dropout layer which only acts on visible APs. Following the augmentation of queries, they are then fed to the multi-head attention layer which has a total of 5 heads (NH) and a head size of 50 (HS). For simplicity of model design, the head size is used as the size of all linear projections within the multi-head attention layer. The output from the multi-head attention layer is fed to a stack of feed-forward fully connected or dense neural network layers with an interleaved dropout of 0.10. We employ the *ReLu* activation function across all layers, except for the output layer which uses *Softmax*. The length of the output layer is set to the number of unique RPs for the given floorplan. With the aforementioned setup, our multi-head attention model design has approximately 111K trainable parameters.

## VI. EXPERIMENTS

### A. Experimental Setup

We evaluated the effectiveness of *ANVIL* across five large indoor paths based on our own measurement across multiple buildings. The next subsection summarizes the approaches from prior work that we compare against, in our experiments.

*1) Fingerprint Benchmark Path Suite*

We evaluate the device invariance of *ANVIL* against four fingerprinting-based indoor localization frameworks from prior work across a benchmark suite containing five indoor paths in different buildings around the Colorado State University campus. *We plan to open-source and release this dataset to benefit the IPIN and broader indoor localization research community.* Figure 6 depicts the floorplans, with each fingerprinted location or reference point (RP) denoted by a yellow dot. We chose a granularity of 1-meter for our experiments (distance between each yellow dot) which we believe is sufficient for the purpose of localizing humans.

The path lengths varied between 60 to 80 meters. Each path was selected due to its salient features that may impact indoor localization. The *Classroom* floorplan is part of one of the oldest buildings on campus that is constructed from wood and concrete. This path is surrounded by a combination of labs that hold heavy metallic equipment as well as large classrooms with open areas. A total of 81 unique WiFi APs were visible on this path. The *Auditorium* and *Library* floorplans are part of relatively new buildings on campus that have a mix of metal

![Figure 7 heatmap table]

Figure 7. Mean localization errors for prior works compared to proposed work (ANVIL) across all combinations of smartphones in the benchmark suite.

and wooden structures with open study areas and bookshelves. We observed 130 and 300 unique APs on the Auditorium and Library floorplans, respectively.

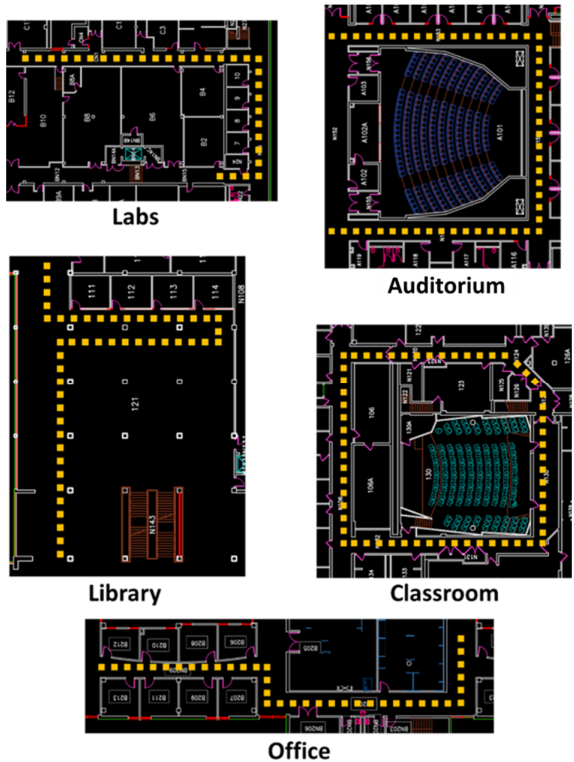

Figure 6: Indoor floorplans within the benchmark suite. Reference points are annotated with yellow boxes.

The *Office* path is on the second floor of an Engineering building that is surrounded by small offices and covered by 180 APs overall. The *Labs* path is in the Engineering basement and is surrounded by labs consisting of a sizable amount of electronic and mechanical equipment with about 120 visible APs. Both the Office and Labs paths have large quantities of metal and electronics that lead to noisy WiFi fingerprints and can hinder indoor localization efforts. A total of 8 fingerprints per RP are used for training *ANVIL* in the offline phase and 2 fingerprints per RP are used in the online phase. We employ six smartphones from distinct vendors to capture the WiFi fingerprints across the five floorplans shown in figure 6. The specifications of these smartphones are listed under table I.

Table I: Details of smartphones used in experiments.

| Smartphone | Chipset | Android version |
|---|---|---|
| OnePlus 3 (OP3) | Snapdragon 820 | 8.0 |
| LG V20 (LG) | Snapdragon 820 | 7.0 |
| Moto Z2 (MOTO) | Snapdragon 835 | 8.0 |
| Samsung S7 (SS7) | Snapdragon 820 | 7.0 |
| HTC U11 (HTC) | Snapdragon 635 | 8.0 |
| BLU Vivo 8 (BLU) | MediaTech Helio P10 | 7.0 |

*2) Comparison with Prior Work*

We identified four state-of-the-art prior works to compare against our proposed *ANVIL* framework. The first work, *LearnLoc* [5] employs the K-Nearest-Neighbor (KNN) approach. It is a lightweight non-parametric approach that employs a Euclidean distance-based metric to match fingerprints. Interestingly, the authors of [27] performed an error analysis to understand the merits of distance calculations. The primary focus was to create a comparison of Euclidean versus Manhattan distance-based metrics to match fingerprints. The work shows evidence of Euclidean distance-based metric to match fingerprints being more accurate. These works are incognizant of device heterogeneity and thus, Euclidean-based *LearnLoc* [5] serves as a one of the motivations for our proposed work. The second work, *AdTrain* [17], is a deep learning-based approach that achieves device invariance through the addition of noise at the input and output labels. This creates an adversarial training scenario where the feed-forward neural network-based model attempts to converge in the presence of high-noise (adversity). The third work, *SHERPA* [28], is similar to the KNN-based approach in [2], [5] but is enhanced to withstand variations across devices in the offline and online phases of fingerprinting-based indoor localization. *SHERPA* achieves variation resilience by employing Pearson's correlation as a distance metric to match the overall pattern of the fingerprints instead of a Euclidian distance-based approach in [5]. Lastly, the fourth work employs Stacked Auto-Encoders (*SAEs*) [16] designed to sustain stable localization accuracy in the presence of device heterogeneity across the offline and online phases. The authors of *SAE* propose using an ensemble of stacked autoencoders (denoising autoencoders) to overcome the variation in RSSI fingerprinting across different devices. A Gaussian process classifier with a radial basis kernel is then employed to produce the final location of the user.

### B. Experimental Results

#### 1) Accuracy Comparsion on Benchmark Suite

A color-coded tabular compilation of the mean localization accuracy in meters across all floorplans in the benchmark suite using every combination of offline (training) and online (testing) devices is presented in figure 7. Each row-group in figure 7 captures the mean indoor localization error in meters for a single floorplan. The device abbreviations on the vertical axes indicate the smartphones used in the offline phase, whereas the devices listed on the horizontal axes indicate smartphones employed by a user in the online phase. The observed localization errors are color coded from green (lower/better) to red (higher/worse) per floorplan.

From figure 7, we immediately observe that the general performance of all five localization frameworks varies across various floorplans. The possible explanation for such an observation could be the differences in path length, overall visibility of WiFi APs, path shape, and other environmental factors. The variances of results across different floorplans also highlights the importance of considering a diverse set of environments when evaluating fingerprinting-based indoor localization frameworks. When making observations about the performance of the different localization frameworks we note that across the whole benchmark suite, *LearnLoc* (with the exception of *SAE*, discussed later) delivers the least stability in the presence of device heterogeneity. The simple introduction of Pearson correlation-based pattern-matching metric through *SHERPA* greatly improves the overall localization accuracy. However, there are some offline-online combinations of smartphones where *SHERPA* is not very effective. For example, in the case when the LG-BLU (offline-online) is used in the Auditorium floorplan, *SHERPA* is not very effective. Based on our empirical analysis, the BLU device exhibits high noise across fingerprints of the same location. Traditional machine learning based classifiers, unlike their deep-learning counterparts, are unable to withstand such noise in fingerprints. In contrast, *AdTrain* and *ANVIL* are both relatively resilient to such noisy combinations of devices which can be attributed to the benefits from adversarial training and FASt layer for fingerprint augmentation, respectively.

From figure 7, we also note that *SAE* produces stable localization performance across device combinations i.e., different offline-online smartphone combinations produce similar localization error. However, the localization errors by themselves are extremely large. Due to such poor performance of *SAE*, we no longer consider it in the subsequent experiments in this paper. *AdTrain* shows the best resilience to device heterogeneity second only to our *ANVIL* framework. We suspect that the simplistic input and label noise induction plays a critical role in the success of *AdTrain*. An interesting observation is that the approaches proposed by the authors in *AdTrain* appear to be orthogonal to the ones proposed by us for the *ANVIL* framework. While *ANVIL* outperforms *AdTrain*, its performance could be further improved by combining the approaches of both frameworks.

Even though figure 7 presents a detailed view of the pair-wise evaluation of various offline-online devices, it is difficult to derive generalized conclusions for the performance of individual frameworks. We therefore capture the average localization errors and associated standard deviations of all offline-online device pairs in figure 8. From figure 8, we note that as compared to *LearnLoc*, both *AdTrain* and *SHERPA* produce better results, with the exception of *AdTrain* in the Classroom and Labs floorplan. In contrast, *ANVIL* consistently produces better results across all floorplans and is more stable (lower standard deviation). *ANVIL* is able to perform up to ~30% better as compared to *SHERPA* and up to ~35% better than *Adtrain* on the Labs and the Office paths.

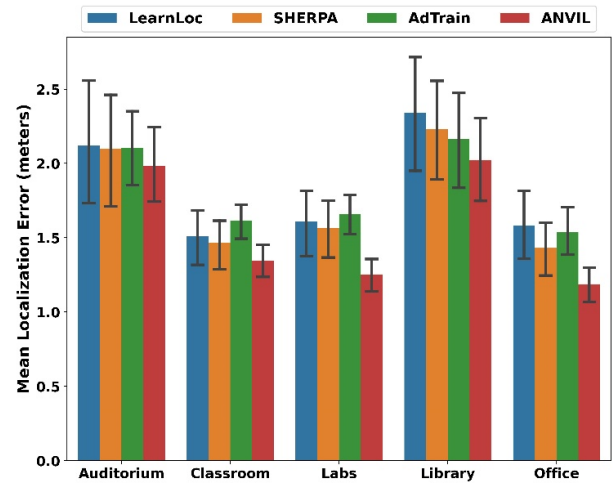

Figure 8. The localization performance of *ANVIL* as compared to state-of-the-art previous works.

In general, based on our experimental analysis in this subsection, we conclude that our proposed *ANVIL* framework outperforms all previous state-of-the-art frameworks considered in this paper, and delivers superior localization accuracy (low error) and stability (similar error across smartphone pairs via low standard deviation) across a diverse set of smartphones and benchmark floorplans.

#### 2) Genralizability of FASt Layer

Within the *ANVIL* framework, we proposed multi-head attention and the FASt layer as two major contributions that play a significant role in strengthening its device invariance. To evaluate the importance of each of these contributions individually and highlight the generalizability of the FASt layer, we apply it to previous works that employ feed-forward FF-DNNs [9] and CNNs (CNNLOC) [6] for fingerprinting-based indoor localization. Our analysis considers two versions of these frameworks: one without the FASt layer (*FF-DNN*,

*CNNLOC*) and one with the FASt layer (*FF-DNN+FASt*, *CNNLOC+FASt*). We also evaluate two versions of *ANVIL*, one with FASt (*ANVIL*) and the other without (*ANVIL+NoFASt*). For the sake of brevity results are averaged across all offline and online phase devices. The results for this analysis are presented in figure 9.

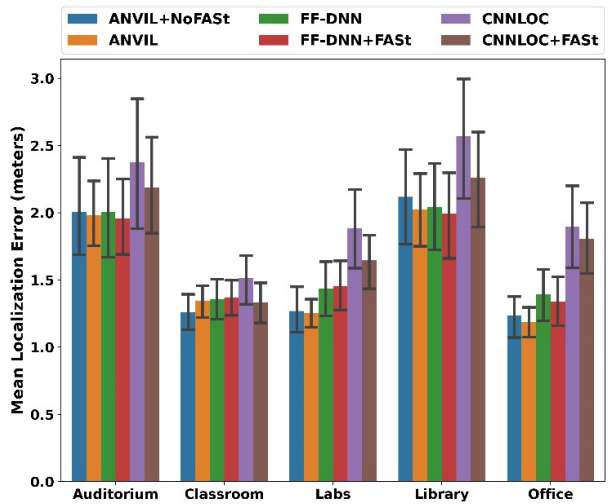

Figure 9. Localization error in meters for *ANVIL* and previous heterogeneity incognizant frameworks with and without the FASt layer.

From figure 9, we note that *CNNLOC*, which is heterogeneity incognizant, produces the highest localization error. It is possible that *CNNLOC* tends to overfit the pattern present in the offline phase device. The *CNNLOC+FASt* variant shows improvement but is still worse than most other frameworks across all paths. Surprisingly *FF-DNN* delivers stronger invariance to device heterogeneity. This can be attributed to the fact that the *FF-DNN* based approach is unable to overfit to the input. The *FF-DNN+FASt* approach offers some limited benefits on Auditorium, Library and the Office paths. Across most of the paths evaluated in figure 9, the *ANVIL* and *ANVIL+NoFASt* produce the best results. The FASt layer when applied to *ANVIL* and also other frameworks leads to improvement in localization accuracy with the exception of the Classroom path. Notably, the Classroom path generally exhibits the least levels of degradation in localization quality as seen in figures 7, 8, and 9. Given that most frameworks achieve ~1.5 meters of average accuracy on the Classroom path, there is very limited room for improvement on that path.

In summary, based on the observations from figure 9, our evaluations strongly highlight the role of 1) the domain adapted fingerprint augmentation (FASt), and 2) multi-head attention, as proposed in *ANVIL*, for the successful realization of device invariant indoor localization. We believe that these two components can be broadly applicable to many other deep learning based indoor localization frameworks, to improve localization accuracy across heterogeneous devices.

VII. CONCLUSION

In this paper, we presented a novel framework called *ANVIL* that combines smart fingerprint augmentation as a layer together with multi-head attention in a neural network, for device invariant indoor localization. We evaluated WiFi RSSI fingerprints from smartphones of different vendors and made key empirical observations that informed our fingerprint augmentation strategy. Our proposed framework was evaluated against several contemporary indoor localization frameworks using six different smartphones across five diverse indoor environments. Through the evaluations, we deduced that *ANVIL* delivers considerable resiliency to device heterogeneity and provides up to 35% better performance compared to the previous works. Our ongoing work is focusing on exploring other attention-mechanisms and incorporating convolutional algorithms into *ANVIL*.